\documentclass[
preprint,
aps
]{revtex4}
\usepackage{graphicx,bm}

\newcommand{\refp}[1]{(\ref{#1})}

\newcommand{\gmbetapara}{\langle \gamma \beta_\parallel \rangle}
\newcommand{\betapara}{\langle \beta_\parallel \rangle}
\newcommand{\IA}{I_A}
\newcommand{\IP}{I_P}
\newcommand{\JP}{J_P}
\newcommand{\rSat}{\tilde{r}}
\newcommand{\rA}{\tilde{r}_A}
\newcommand{\rP}{\tilde{r}_P}
\newcommand{\Bmax}{B_\mathrm{max}}
\newcommand{\lmdlin}{\lambda_0}
\newcommand{\rlin}{r_0}

\begin{document}

%%%%%%%  Title
\title{Saturation mechanism of the Weibel instability
in weakly magnetized plasmas}%
\author{Tsunehiko N. \surname{Kato}}%
\email{katoutn@cc.nao.ac.jp}
\affiliation{
National Astronomical Observatory of Japan,
2-21-1 Osawa, Mitaka, Tokyo 181-8588, Japan 
}
%\pacs{52.35.Qz, 52.65.Rr, 95.30.Qd}

\date{27 June, 2005}
%\date{\today}

%%%%%%%  Abstract
\begin{abstract}
The saturation mechanism of the Weibel instability is investigated theoretically
by considering the evolution of currents in numerous cylindrical beams
that are generated in the initial stage of the instability.
Based on a physical model of the beams,
%,which is applicable to relativistic plasmas as well as nonrelativistic plasmas.
it is shown that
the magnetic field strength attains a maximum value
when the currents in the beams evolve into the Alfv\'{e}n current
and that there exist two saturation regimes.
The theoretical prediction of the magnetic field strength at saturation is
in good agreement with the results of
two-dimensional particle-in-cell simulations
for a wide range of initial anisotropy.
%for both electron-positron and electron-proton plasmas.
\end{abstract}
%%%%%%%

\maketitle

%%%%%%%
%\section{Introduction}
%%%%%%%
Collisionless plasmas with anisotropic velocity distributions
drive the Weibel instability \cite{Weibel59, Fried59},
thereby generating magnetic fields.
Recently,
this instability in astrophysical as well as laboratory plasmas
has attracted considerable attention.
For example,
it is considered that
this instability can be driven in strong collisionless shock waves
associated with various astrophysical phenomena,
e.g.,
pulsar winds \cite{Hoshino98,Kazimura98},
gamma-ray bursts and/or their afterglows \cite{Medvedev99},
or gravitational collapse of
large-scale structures in the universe \cite{Schlickeiser03}.
It can also be driven
if temperature gradients are present in plasmas \cite{Okabe03}.
The magnetic field generated by the instability
is responsible for 
synchrotron and/or ``jitter'' radiation
from the existing high-energy particles\cite{Medvedev00}.
Furthermore,
such magnetic fields
can provide an effective scattering mechanism
for charged particles.
For example,
it would affect
the dissipation process of collisionless shock waves,
efficiency of the Fermi acceleration,
or heat conductivity by charged particles.
In all these cases,
the amplitude of the magnetic field is of primary importance.

The magnetic field strength
attains the maximum value at the saturation of the instability.
Several conditions for saturation
have been proposed \cite{Davidson72, Yang94, Medvedev99}.
However, these conditions may be unsatisfactory
because the role of current was not treated explicitly in these studies.
Since the instability enters the nonlinear regime before saturation,
the currents and fields must be considered self-consistently.
As shown by numerical simulations,
the currents are carried by numerous cylindrical beams \cite{Lee73, Silva03,Frederiksen04}.
Therefore,
the physics of saturation will be elucidated by examining the evolution of such beams.

In this letter, the nonlinear saturation mechanism of the Weibel instability,
i.e., that of the transverse modes of filamentation instability,
is investigated with a physical model of the beams.
The theoretical analysis and simulations are performed
within a two-dimensional framework.
(Therefore, in three dimensions, the results might be modified by other effects.)
The method is applicable to
electron-proton as well as electron-positron plasmas.
However,
for electron-proton plasmas,
only the saturation due to the electron currents is considered.
Although proton currents later cause the \textit{second} saturation\cite{Frederiksen04},
it is beyond the scope of the present study
and will be investigated in the future.
In the following analysis,
protons are treated as a charge neutralizing background.

%%%%%%%
\textit{Theoretical model of saturation---}
%\paragraph{Theoretical model of saturation ---}
%%%%%%%
In this section,
the saturation mechanism
is modeled
by using Gaussian units and the following notation:
speed of light, $c$;
velocity of a particle normalized by the speed of light, $\bm{\beta}$;
Lorentz factor of the particle, $\gamma \equiv (1-\bm\beta^2)^{-1/2}$;
electron mass, $m_e$;
electron charge, $-e$;
proton mass, $m_p$;
mean electron number density, $n_e$;
electron plasma frequency,
$\omega_{pe} \equiv (4\pi n_e e^2/m_e)^{1/2}$;
electron skin depth,
$l_0 \equiv c/\omega_{pe} \approx 5.3\times 10^5\ n_e^{-1/2}\ [\mathrm{cm}]$.

The initial condition is set as follows.
The plasma is initially
uniform and unmagnetized (or weakly magnetized)
with an anisotropic velocity distribution.
The initial velocity distributions of the particles are axisymmetric,
and the velocity dispersion along the $z$-axis
is larger than those along the other axes.
For electron-positron plasmas,
the distribution functions are the same
for both species.
For electron-proton plasmas,
the distribution function of only the electrons is considered.
Such a distribution function
would be plausible for two-stream like conditions
\cite{Lee73, Kazimura98, Califano98, Medvedev99, Silva03, Schlickeiser03, Frederiksen04}.
%The opposite anisotropy cases
%\cite{Weibel59, Davidson72, Yang94, Hoshino98}
%are deferred to future work.

The basic physical mechanism of the instability is described as follows.
In the initial stage of the instability,
the charged particles in the plasma
are deflected by small magnetic fluctuations.
In the present condition,
most particles have larger velocities along the $z$-axis.
Since the particles carrying current toward the $+z$ direction
and $-z$ direction
are deflected in opposite directions,
they are separated into different regions,
thereby producing net currents in the plasma
that generate magnetic fields\cite{Medvedev99}.
As the amplification of the magnetic fields increases,
the distance between the two populations of particles increases,
and the currents are amplified,
and vice-versa.
Thus,
%this process acts as an instability;
both currents and magnetic fields increase exponentially.
In this evolution,
it is important for the instability to enter the nonlinear regime before saturation.
As revealed by two- or three-dimensional simulations \cite{Lee73, Silva03,Frederiksen04},
many isolated cylindrical beams (or current filaments)
are formed in the plasma long before saturation;
each beam carries a net current and generates a magnetic field around itself.
Since two beams with currents in the same direction attract each other
and tend to coalesce into a larger beam,
the beams grow with time \cite{Lee73, Silva03}
even before saturation.

Based on these observations,
we model the nonlinear evolution of the instability
from the ``initial condition'' in which the plasma consists of many cylindrical beams
to saturation.
Initially,
the radius and magnitude of current in the beams are almost uniform
because the plasma is homogeneous.
Each beam carries a net current in either the $+z$ or $-z$ direction.
Subsequently,
both the current and magnetic field in each beam
increase exponentially due to particle separation in the beam.
The radii of the beams also increase due to the coalescing process,
although it is considerably slower than the increase in current and magnetic field.
However,
they are still considered to be uniform %at any point of time
because they essentially evolve equally.
Therefore,
the plasma can be modeled as an ensemble of uniform cylindrical beams
with the same radius $r$ and magnitude of current $I$.
(A similar model was recently considered in Ref.~\cite{Medvedev05}
for investigating the evolution after saturation.)
By assuming a uniform current density $J$,
where $I = \pi r^2 J$,
the root-mean-square value of the magnetic field strength within a cylindrical beam
is calculated using the Amp\`{e}re's law as follows:
\begin{equation}
	B = \sqrt{2} \pi r  J / c = \sqrt{2} I / (rc).
	\label{eq:B}
\end{equation}
Although in practice, the magnetic field is affected by neighboring beams,
their net effect will be low and the order of the magnetic field strength
in the beam will be given by Eq.~\refp{eq:B}.
Thus,
we assume that the magnetic field \refp{eq:B} in each beam
is essential to the saturation process
and
those generated by the neighboring beams are important only in the coalescing process.

On the basis of this model,
we consider the evolution of one of the beams.
The exponential increase in current and magnetic field
ceases when
the magnitude of the current is equal to the smaller of the following two upper limits.
One limit is the Alfv\'{e}n current \cite{Alfven39},
which occurs due to the self-generated magnetic field;
it is expressed in Gaussian units by
\begin{equation}
	\IA = I_0 \gmbetapara,
\end{equation}
where $I_0 \equiv m_e c^3 / e\ (\approx 17000 \mathrm{[A]})$,
$\beta_\parallel$ is the \textit{magnitude} of
the $z$ component of $\bm\beta$,
and the angle brackets $\langle\ \rangle$ denote an average taken over the beam volume.
(This limit was originally derived for a monoenergetic single-directed particle beam.
In this case,
the averaged value of $\gamma\beta_\parallel$ is used
instead of that in the monoenergetic cases
because the energy of particles is distributed.)
We can apply this limit even when
the separation process is in progress and
a fraction of particles
carry current in the opposite direction
because the self-generated magnetic field
is associated with the \textit{net} current $I$,
and it always ensures that $I$ does not exceed $\IA$.
The other limit is the maximum current $\IP$
that can be carried by all the particles within the beam:
\begin{equation}
	\IP = \pi r^2 \JP,
	\quad
	\mathrm{where}
	\quad
	\JP \equiv e \mu n_e \betapara c
\end{equation}
with
$\mu = 1$ for electron-proton plasmas
and
$\mu = 2$ for electron-positron plasmas.
This limit is reached when
the separation of the current density is completed
before $I$ equals $\IA$.
Since the distribution of particles in the beam
approaches an isotropic distribution
depending on the evolution of the instability,
we must consider this effect;
it is expressed by a factor $\chi$ as follows:
\begin{equation}
	\betapara = \chi \betapara_0,
	\qquad
	\gmbetapara = \chi \gmbetapara_0,
	\label{eq:chi}
\end{equation}
where
subscript ``0'' denotes an initial value.
Since a complete isotropization of the strong initial anisotropy
yields the lower limit $\chi = 1/3$,
we observe that $1/3 < \chi < 1$.

After the exponential growth ceases,
the beam evolves through the coalescing process.
In the former case ($\IA < \IP$),
which is termed \textit{the Alfv\'{e}n limit},
once the current equals the Alfv\'{e}n current $\IA$,
it no longer increases
and it retains the value of $\IA$,
while the radius continues to increase.
Therefore,
from Eq.~\refp{eq:B},
the magnetic field decreases monotonically thereafter.
In the latter case $(\IA > \IP)$,
which is termed \textit{the particle limit},
current gradually increases
until it reaches the Alfv\'{e}n current;
during this period,
it increases with the radius
while satisfying the condition $I=\pi r^2 \JP$,
and the magnetic field also increases.
After the current equals the Alfv\'{e}n current,
the magnetic field decreases monotonically,
as in the Alfv\'{e}n limit.

Therefore,
for both limits,
the magnetic field strength becomes maximum
when the current evolves into the Alfv\'{e}n current.
From Eq.~\refp{eq:B},
the maximum value $\Bmax$ is expressed as follows:
\begin{equation}
	\Bmax	= \sqrt{2} \IA / (\rSat c)
			= \sqrt{2}\gmbetapara (\rSat/l_0)^{-1} B_*,
	\label{eq:Bmax}
\end{equation}
where
$\rSat$ is the beam radius at saturation
and $B_*$ is the magnetic field strength given by
$B_* \equiv c (4\pi n_e m_e)^{1/2} \approx 3.2 \times 10^{-3} n_e^{1/2}\ \mathrm{[G]}$.
In the Alfv\'{e}n limit,
the radius at saturation $\rA$
is related to the initial radius $\rlin$
or the wavelength of the most unstable mode
in the linear theory $\lmdlin$.
Here,
it is simply expressed using two factors $\alpha$ and $\zeta$ as follows:
\begin{equation}
	\rA = \alpha \rlin
	= \alpha \zeta \lmdlin.
	\label{eq:r_A}
\end{equation}
Thus,
the magnetic field at saturation is given by
\begin{equation}
	\Bmax
	 = \sqrt{2} c_A \gmbetapara_0 \left( \lmdlin / l_0 \right)^{-1} B_*,
 	 \label{eq:Bmax_A}
\end{equation}
where $c_A \equiv \chi_A/(\alpha\zeta)$,
and
$\chi_A$ denotes the effect of isotropization at saturation
[see Eq.~\refp{eq:chi}].
In the particle limit,
the radius at saturation $\rP$
is determined
from the condition $\pi \rP^2 \JP = \IA$
as follows:
\begin{equation}
	\rP
	 = (\IA / \pi\JP)^{1/2}
	 = 2 l_0 \left( \gmbetapara_0 /  \mu \betapara_0 \right)^{1/2}
	\label{eq:r_P}
\end{equation}
which is on the order of $l_0$
for nonrelativistic cases.
Using this radius,
the maximum magnetic field strength is given by
\begin{equation}
	\Bmax
	 = \chi_P \left( \mu \gmbetapara_0 \betapara_0 / 2 \right)^{1/2} \ B_*,
	 \label{eq:Bmax_P}
\end{equation}
where $\chi_P$ is the isotropization factor at saturation.
%The condition to determine the transition point
%is given by $\IA = \pi\rA^2 \JP$, or $\rA = \rP$.
As mentioned earlier,
the values of $\Bmax$
in Eqs.~\refp{eq:Bmax_A} and \refp{eq:Bmax_P}
can be regarded as
typical magnetic fields of the entire plasma at saturation
in the respective cases.
The two parameters ($c_A$ and $\chi_P$) in these expressions
are expected to be of order unity.

The ratio of the maximum magnetic energy density
to the initial particle kinetic energy density, $\eta$,
is written as
$\eta = (\Bmax/B_*)^2 \ [2\mu(\langle \gamma \rangle-1)]^{-1}$.
(It should be noted that for electron-proton plasmas,
the kinetic energy of protons is \textit{not} considered in this equation.)
For the particle limit regime,
$\eta$ yields a subequipartition value.
In particular,
$\eta \sim \chi_P^2/2$ for nonrelativistic plasmas
and $\eta \sim \chi_P^2/4$ for ultrarelativistic plasmas.
The conclusion
that $\eta$ yields a subequipartition value
for strongly anisotropic cases
is in agreement with that of previous simulations
\cite{Davidson72, Yang94, Silva03, Califano98, Kazimura98}.

%%%%%%%
\textit{Comparison with numerical simulations ---}
%\paragraph{Comparison with numerical simulations ---}
%%%%%%%
The results of the developed analytical model
are compared
with those of numerical simulations
for electron-positron
and electron-proton plasmas ($m_p/m_e=1836$).
The simulation code used is a relativistic, electromagnetic, particle-in-cell code
with two spatial and three velocity dimensions.
This code was developed based on a general description by Ref. \cite{Birdsall}.
The $x$-$y$ plane perpendicular to the $z$-axis
is considered to be the simulation plane.
The initial particle distribution
is expressed in terms of the normalized momentum $\mathbf{u} \equiv \gamma \bm{\beta}$
that is common for all the species.
Each component of $\mathbf{u}$ obeys the Gaussian distribution
with a standard deviation of
$\sigma_\parallel$ for the $z$ component
or $\sigma_\perp$ for the other components.
The simulations were performed 
using a $512 \times 512$ grid with $\sim 50$ particles per cell per species
under periodic boundary conditions
for several values of $\sigma_\parallel$ ranging from $0.12$ to $10$
with fixed $\sigma_\perp = 0.1$.
For each simulation,
the physical size of the simulation box in each direction, $L$,
was considered to be at least $7$ times larger than
the typical beam diameter at saturation,
which was estimated by a preliminary simulation
[e.g., $L=120 l_0$ for $\sigma_\parallel=0.12$ (largest);
$L=30 l_0$ for $\sigma_\parallel=0.6$ (smallest)].
Then,
in each simulation,
the evolution of the magnetic field strength
averaged over the simulation box was monitored
and its maximum value was obtained;
the calculation time $T$ considered was
long enough to obtain the maximum value
[e.g., $T = 1500 \omega_\mathrm{pe}^{-1}$ for $\sigma_\parallel = 0.12$ (longest);
$T = 25 \omega_\mathrm{pe}^{-1}$ for $\sigma_\parallel = 1.1$ (shortest)].
Figure~1 % ~\ref{fig:Bmax}
shows the obtained maximum values denoted by dots
as a function of the initial anisotropy, $\sigma_{\parallel} /\sigma_{\perp} - 1$:
(a) electron-positron plasma and
(b) electron-proton plasma.

The corresponding theoretical results are obtained as follows:
First, in Eqs.~\refp{eq:Bmax_A} and \refp{eq:Bmax_P},
we approximate that
$\gmbetapara_0 = \sigma_\parallel$
and $\betapara_0 = \sigma_\parallel / \tilde{\gamma}$,
where
$\tilde{\gamma} \equiv (1 + \sigma_\parallel^2 + 2 \sigma_\perp^2)^{1/2}$.
Next,
in the present case,
since the linear theory of nonrelativistic weak anisotropic plasmas \cite{Davidson72}
is applicable to the Alfv\'{e}n limit regime,
we obtain
\begin{equation}
	\lmdlin = 2\pi \sqrt{3/\mu} \left[(\sigma_{\parallel} /\sigma_{\perp})^2-1 \right]^{-1/2} \ l_0.
	\label{eq:lmdz}
\end{equation}
(It should be noted that,
when $\sigma_\perp > 1$ or $\sigma_\parallel > 1$,
the relativistic dispersion relation must be employed.)
Finally,
using these expressions,
we obtain
\begin{equation}
	\frac{\Bmax}{B_*} =
		\left\{
			\begin{array}{rc}
				\displaystyle
					\sqrt{\frac{\mu}{6}}
					\frac{c_A\sigma_\parallel}{\pi}
					\left[ (\sigma_{\parallel} /\sigma_{\perp})^2-1 \right]^{1/2}
				& \mbox{(Alfv\'{e}n limit)} \\
				\chi_P \sigma_\parallel \left[\mu/(2\tilde{\gamma})\right]^{1/2}
				& \mbox{(particle limit)}
				.
			\end{array}
		\right.
	\label{eq:Bmax_AP2}
\end{equation}
The results
are plotted in Fig.~1; %\ref{fig:Bmax}:
dotted curves represent the results for the Alfv\'{e}n limit
and solid curves for the particle limit.
The parameters ($c_A$ and $\chi_P$) are assumed to be constants.
Their values are taken as
$(c_A, \chi_P) = (1.2, 0.5)$ in (a)
and
$(1.0, 0.5)$ in (b)
to match with the simulation results;
they are of order unity, as expected.
The transition anisotropy is given by
$(\sigma_\parallel/\sigma_\perp)_c \approx [1 + 3\pi^2 (\chi_P / c_A)^2]^{1/2}$,
and
$(\sigma_\parallel/\sigma_\perp)_c \approx 2.5$ for (a) and $\approx 2.9$ for (b).
%%%
%\begin{figure}[htbp]
%	\includegraphics{fig1.eps}
%	\caption{
%		\label{fig:Bmax}
%Maximum magnetic field strength $B_\mathrm{max}$
%for $\sigma_\perp=0.1$
%as a function of the initial anisotropy, $\sigma_{\parallel} /\sigma_{\perp} - 1$:
%\textbf{(a)} electron-positron plasma,
%\textbf{(b)} electron-proton plasma.
%The dots are results from numerical simulations.
%The dashed curves and the solid curves represent the theoretical results 
%for the Alfv\'{e}n limit
%and the particle limit, respectively [see Eq.~\refp{eq:Bmax_AP2}].
%This shows good agreement between the prediction of the proposed theory
%and simulation results in each saturation regime.
%	}
%\end{figure}
%%%
These curves are in good agreement with
the simulation results
in their respective regions.
It is evident that
there are two saturation regimes.
We also observe that
the assumption of the constancy of $c_A$ and $\chi_P$
holds true over a wide range of anisotropy in both the figures,
although the theoretical curves deviate slightly from
those of the simulation results in the weak anisotropy side.

Other comparisons are shown below for electron-positron plasmas.
Using Eq.~\refp{eq:B},
we can indirectly estimate $r$
from the simulation results:
\begin{equation}
	r =  (2\pi\sqrt{2})^{-1} (B/B_*) (S/l_0^2) (I_\mathrm{tot}/I_0)^{-1} l_0,
	\label{eq:r_max}
\end{equation}
where $S$ is the area of the simulation box
and $I_\mathrm{tot}$ is the total current in one direction along the $z$-axis.
Figure~2(a) %\ref{fig:r_max}(a)
compares the radius at saturation obtained from this equation
with those of the model
($\tilde{r}_A$ [Eqs.~\refp{eq:r_A} and \refp{eq:lmdz} with $\alpha \zeta = 0.7$]
and $\tilde{r}_P$ [Eq.~\refp{eq:r_P}]).
Figure~2(b) %\ref{fig:r_max}(b)
shows
the current per beam at saturation
estimated using the radius of Eq.~\refp{eq:r_max}
normalized by the Alfv\'{e}n current $\IA = I_0 \sigma_\parallel$,
which does \textit{not} include the isotropization effect.
Hence, the normalized values
are not expected to be unity but
to be close to the isotropization factor;
$\chi_A = c_A \alpha\zeta = 0.84$ for the Alfv\'{e}n limit, and
$\chi_P = 0.5$ for the particle limit (shown by dotted lines).
We observe that the model is consistent with the simulations.
%%%
%\begin{figure}[htbp]
%	\includegraphics{fig2.eps}
%	\caption{
%		\label{fig:r_max}
%\textbf{(a)} Typical beam radius at saturation
%for electron-positron plasma
%as a function of the initial anisotropy with $\sigma_\perp=0.1$.
%The dots show the radii of Eq.~\refp{eq:r_max} evaluated using the simulation results.
%The dashed curve represents $\tilde{r}_A$ [see Eq.~\refp{eq:r_A}].
%The solid curve shows $1.2\ \tilde{r}_P$ [see Eq.~\refp{eq:r_P}].
%\textbf{(b)} Current per beam at saturation obtained from the simulation results
%normalized by the Alfv\'{e}n current ($\IA = I_0 \sigma_\parallel$) .
%The expected values are shown by the dotted lines (see text).
%It is seen that
%the theoretical model is consistent with the numerical simulations.
%	}
%\end{figure}
%%%
Figure~3 %\ref{fig:Jz}
shows the current density at saturation
obtained from the simulations
for three cases: (a) Alfv\'{e}n limit regime ($\sigma_\parallel = 0.15, \chi=0.84$),
(b) particle limit regime ($\sigma_\parallel = 3.1, \chi=0.5$),
and (c) the transition point ($\sigma_\parallel = 0.3, \chi=0.5$).
The current density is normalized by $\JP$, which includes the isotropization effect.
It is evident that saturation occurred when $|J_z| \approx \JP$ for (b) and (c),
while $|J_z| \ll \JP$ for (a).
It should be noted that
in some regions of (b) and (c),
the current density exceeds $\JP$.
This occurs because the current-carrying beams are pinched.
In any case,
the typical magnetic field can still be estimated from Eq.~\refp{eq:B}
because it is mainly determined
by the total current in a beam,
which approximates to the Alfv\'{e}n current at saturation
irrespective of whether the beam is pinched or not.
%%%
%\begin{figure}[htbp]
%	\includegraphics{fig3.eps}
%	\caption{
%		\label{fig:Jz}
%Contour plots of current density at saturation, $J_z$, normalized by $J_P$,
%for $\sigma_\perp = 0.1$:
%\textbf{(a)} Alfv\'{e}n limit regime ($\sigma_\parallel = 0.15$),
%\textbf{(b)} Particle limit regime ($\sigma_\parallel = 3.1$),
%and \textbf{(c)} the transition anisotropy ($\sigma_\parallel = 0.3$).
%The horizontal and vertical axis are $x$- and $y$-axis, respectively,
%in unit of $l_0$.
%The dotted curves show the levels of $J_z/J_P = \pm 0.1$ in (a),
%and those of $J_z/J_P = \pm 1$ in (b) and (c).
%The solid curves show that of $J_z = 0$ in all figures.
%We see that
%the saturation occurs when $|J_z| <J_P$ in (a)
%and when $|J_z| \sim J_P$ in (b) and (c),
%as predicted by the proposed theory.
%This figure also confirms that the beams are approximately uniform
%in radius and in current for all cases,
%which is assumed in the theoretical model.
%	}
%\end{figure}
%%%

%%%%%%%
\textit{Discussion ---}
%\paragraph{Discussion ---}
%%%%%%%
Even in the presence of a background magnetic field,
the proposed model is applicable
if the time taken for saturation
is shorter than the deflection time due to the background magnetic field.
Otherwise,
the problem of the magnetized Weibel instability
or the whistler instability for electron-proton plasmas \cite{Ossakow72}
should be considered.

It is shown that
the proposed model is consistent with several conditions
obtained previously.
Medvedev and Loeb \cite{Medvedev99}
proposed that
saturation occurs
when the effective Larmor radius $r_L(B)$
becomes comparable to
the most unstable wavelength in linear theory, $\lmdlin$.
If $\lmdlin$ is replaced by the typical beam radius at saturation, $\rSat$,
we obtain the condition of $r_L(\Bmax) \approx \rSat$,
which is qualitatively
equivalent to condition \refp{eq:Bmax}.
Califano \textit{et al}.\ \cite{Califano98} also found that
saturation occurs when $r_L(B) \approx l_0$.
This result is in agreement with the result of the particle limit
in Eq.~\refp{eq:r_P}.

In electron-proton plasmas,
if the initial velocity distribution of the protons is the same as that of electrons,
the upper limits for the proton currents
will be given by $I'_A = (m_p/m_e) \IA$ and $I'_P = \IP$ with $\mu=1$.
Thus,
for the particle limit,
both $\rP$ and $\Bmax$ would be
$(m_p/m_e)^{1/2}$ times larger than those of the electron currents.
%The Alfv\'{e}n limit cannot be discussed here
%because how $\rA$ is determined
%has not yet been elucidated for proton currents.

In conclusion,
the magnetic field generated by the Weibel instability
saturates when the currents in the beams
evolve into the Alfv\'{e}n current;
there are two saturation regimes:
the Alfv\'{e}n limit and the particle limit.
The beam model proposed in this letter
provides a good estimate
of the magnetic field strength at saturation.
This model will also be useful to consider the evolution of
magnetic fields even after saturation.

%%%%%%%  Acknowledgments
\begin{acknowledgments}
I would like to thank M.~Hattori, Y.~Fujita, and N.~Okabe
for their helpful discussions.
This research is partly supported
by the Japan Science and Technology Agency.
\end{acknowledgments}

\newpage

\newpage
\begin{figure}[htbp]
	\includegraphics{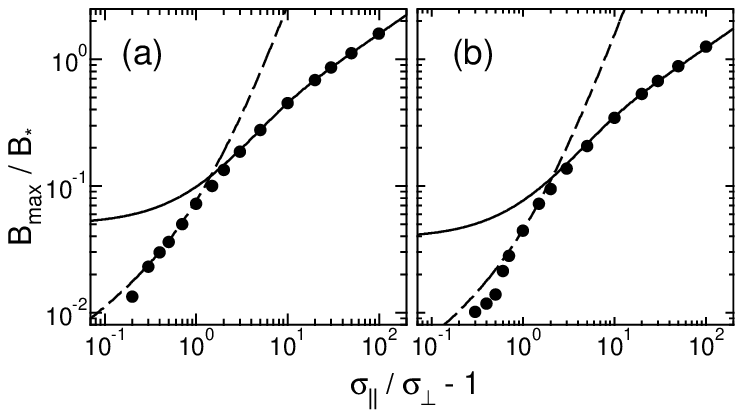}
	\caption{
	Maximum magnetic field strength $B_\mathrm{max}$
	for $\sigma_\perp=0.1$
	as a function of the initial anisotropy, $\sigma_{\parallel} /\sigma_{\perp} - 1$:
	\textbf{(a)} electron-positron plasma, and
	\textbf{(b)} electron-proton plasma.
	Dots denote the results of numerical simulations.
	The dashed and solid curves represent the theoretical results 
	for the Alfv\'{e}n limit
	and particle limit, respectively [see Eq.~\refp{eq:Bmax_AP2}].
	This shows good agreement between the predictions of the proposed theory
	and the simulation results in each saturation regime.
	}
\end{figure}

\newpage
\begin{figure}[htbp]
	\includegraphics{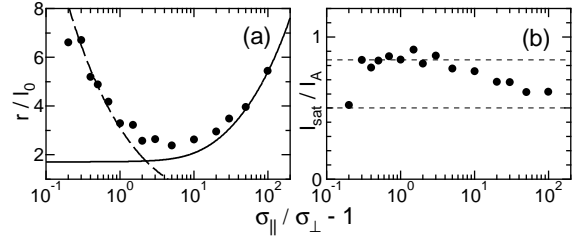}
	\caption{
	\textbf{(a)} Typical beam radius at saturation
	for the electron-positron plasma
	as a function of the initial anisotropy with $\sigma_\perp=0.1$.
	Dots represent the radii of Eq.~\refp{eq:r_max} evaluated using the simulation results.
	The dashed curve represents $\tilde{r}_A$ [see Eq.~\refp{eq:r_A}];
	solid curve $1.2\ \tilde{r}_P$ [see Eq.~\refp{eq:r_P}].
	\textbf{(b)} Current per beam at saturation obtained from the simulation results
	normalized by the Alfv\'{e}n current ($\IA = I_0 \sigma_\parallel$) .
	The expected values are represented by the dotted lines (see text).
	The results of the theoretical model are
	consistent with those of the numerical simulations.
	}
\end{figure}

\newpage
\begin{figure}[htbp]
	\includegraphics{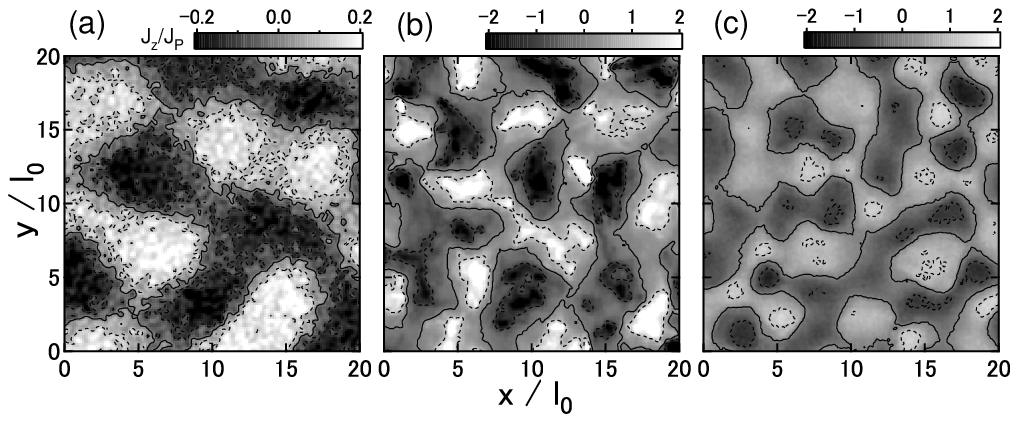}
	\caption{
	Contour plots of current density at saturation, $J_z$, normalized by $J_P$,
	on the $x-y$ plane
	for $\sigma_\perp = 0.1$:
	\textbf{(a)} Alfv\'{e}n limit regime ($\sigma_\parallel = 0.15$),
	\textbf{(b)} particle limit regime ($\sigma_\parallel = 3.1$),
	and \textbf{(c)} the transition anisotropy ($\sigma_\parallel = 0.3$).
	The spatial unit is the electron skin depth $l_0$.
	The dotted curves indicate the levels of $J_z/J_P = \pm 0.1$ in (a),
	and those of $J_z/J_P = \pm 1$ in (b) and (c).
	The solid curves indicate the levels of $J_z = 0$ in all figures.
	We observe that
	saturation occurs when $|J_z| <J_P$ in (a)
	and $|J_z| \sim J_P$ in (b) and (c),
	as predicted by the proposed theory.
	This figure also confirms that the beams are approximately uniform
	in radius and current for all the cases,
	which are assumed in the theoretical model.
	}
\end{figure}


\begin{thebibliography}{}
%
\bibitem{Weibel59} E.~S.~Weibel, Phys. Rev. Lett. \textbf{2}, 83 (1959).
%
\bibitem{Fried59} B.~D.~Fried, Phys. Fluids \textbf{2}, 337 (1959).
%
\bibitem{Hoshino98} M.~Hoshino, in \textit{Neutron stars and pulsars},
%proceedings of the International Conference on Neutron Stars and Pulsars, Tokyo, 
edited by N. Shibazaki \textit{et al}. (Univ. Academy Press, 1998), p.~491
%
\bibitem{Kazimura98} Y.~Kazimura, J.~I.~Sakai, T.~Neubert, and S.~V.~Bulanov,
Astrophys. J. Lett. \textbf{498}, L183 (1998).
%
\bibitem{Medvedev99} M.~V.~Medvedev and A.~Loeb, Astrophys. J. \textbf{526}, 697 (1999).
%
\bibitem{Schlickeiser03} R.~Schlickeiser and P.~K.~Shukla, Astrophys. J. Lett. \textbf{599}, L57 (2003).
%
\bibitem{Okabe03} N.~Okabe and M.~Hattori, Astrophys. J. \textbf{599}, 964 (2003).
%
\bibitem{Medvedev00} M.~V.~Medvedev, Astrophys. J. \textbf{540}, 704 (2000).
%
\bibitem{Davidson72} R.~C.~Davidson, D.~A.~Hammer, I.~Haber, and C.~E.~Wagner,
Phys. Fluids \textbf{15}, 317 (1972).
%
\bibitem{Yang94} T.-Y.~B.~Yang, J.~Arons, and A.~B.~Langdon, Phys. Plasmas \textbf{1}, 3059 (1994).
%
\bibitem{Lee73} R.~Lee and M.~Lampe, Phys. Rev. Lett. \textbf{31}, 1390 (1973).
%
\bibitem{Silva03} L.~O.~Silva, R.~A.~Fonseca, J.~W.~Tonge, J.~M.~Dawson, W.~B.~Mori, and M.~V.~Medvedev,
Astrophys. J. Lett. \textbf{596}, L121, (2003).
%
\bibitem{Frederiksen04} J.~T.~Frederiksen, C.~B.~Hededal, T.~Haugb\o{}lle, and \r{A}.~Nordlund,
Astrophys. J. Lett. \textbf{608}, L13, (2004).
%
\bibitem{Califano98} F.~Califano, F.~Pegoraro, S.~V.~Bulanov, and A.~Mangeney,
Phys. Rev. E \textbf{57}, 7048 (1998).
%
\bibitem{Medvedev05} M.~V.~Medvedev, M.~Fiore, R.~A.~Fonseca, L.~O.~Silva, and W.~B.~Mori,
Astrophys. J. Lett. \textbf{618}, L75, (2005).
%
\bibitem{Alfven39} H.~Alfv\'{e}n, Phys. Rev. \textbf{55}, 425 (1939).
%
\bibitem{Birdsall} C.~K.~Birdsall and A.~B.~Langdon,
\textit{Plasma Physics via Computer Simulation} (IOP Publishing, Bristol, 1991).
%
\bibitem{Ossakow72} S.~L.~Ossakow, I.~Haber, and E.~Ott,
Phys. Fluids \textbf{15}, 1538 (1972).
%
\end{thebibliography}
\end{document}